# Geant4 and beyond: recent progress in precision physics modeling

Matej Batic[2], Marcia Begalli[3], Min Cheol Han[4], Steffen Hauf[5], Gabriela Hoff[6], Chan Hyeong Kim[4], Han Sung Kim[4], Sung Hun Kim[4], Markus Kuster[5], Maria Grazia Pia[1], Paolo Saracco[1], Georg Weidenspointner[5]

[1]*INFN Sezione di Genova, Via Dodecaneso 33, Genova 16146, Italy*
[2]*Sinergise, Ljubljana, Slovenia*
[3]*State University Rio de Janeiro, Rio de Janeiro, Brazil*
[4]*Hanyang University, Seoul, Korea*
[5]*XFEL GmbH, Hamburg, Germany*
[6]*PUCRS, Porto Alegre, Brazil*

## INTRODUCTION

This extended abstract briefly summarizes ongoing research activity on the evaluation and experimental validation of physics methods for photon and electron transport. The analysis includes physics models currently implemented in Geant4 [1,2] as well as modeling methods used in other Monte Carlo codes, or not yet considered in general purpose Monte Carlo simulation systems.

The validation of simulation models is performed with the support of rigorous statistical methods, which involve goodness-of-fit tests followed by categorical analysis. All results are quantitative, and are fully documented.

## PHYSICS MODELS AND THEIR VALIDATION

Each one of the physics modeling domains outlined below is, or will be, addressed in one or more dedicated publications in scholarly journals. The typical length of these papers is 20-30 pages each; due to the large amount of validation results, it is practically impossible to document all of them in this extended abstract. Therefore this extended abstract is intended to provide only an overview of the subjects proposed for presentation at RPSD 2014; it is not meant to replace future detailed publications. It is worthwhile to note that the effect of the accuracy of physics models on the overall accuracy of a simulation outcome cannot be univocally determined a priori, as it depends on the experimental scenario and the simulation observables of interest.

### Photon elastic scattering

The analysis [3] addressed modeling methods based on the form factor approximation and on 2nd order S-matrix calculation [4]. Non-relativistic [5], relativistic [6], and modified [7] form factors were considered. The validation tests involved more than 3000 experimental measurements.

Statistical analysis identified the model based on S-matrix calculation as the most accurate one, although its computational performance is slower than for models implementing the form factor approximation due to more complex bi-dimensional interpolation algorithms.

An example of comparison between simulation models and experimental data is shown in Figure 1.

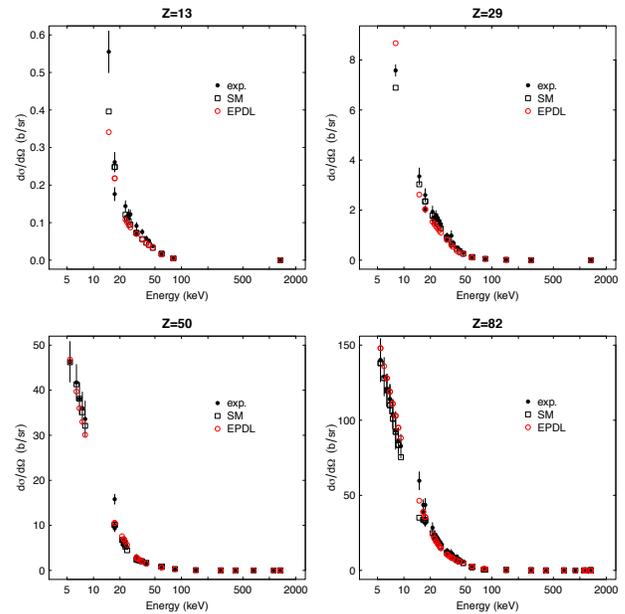

Fig. 1 An example of photon elastic scattering cross section validation: experimental measurements (black symbols) and a selection of differential cross section calculations considered in the validation analysis.

### Compton scattering

Implementations of Compton scattering in Geant4 low energy and standard packages were subject to validation: they are based on EPDL [8], Penelope [9] original implementation and an empirical fit to cross sections calculated by Hubbell [10] and Storm and Israel [11], respectively. Statistical analysis showed that all the considered models reproduce experimental data equivalently well.

The validation of differential cross sections concerned the Klein-Nishina [12] cross section formula and models based on a variety of scattering functions [13,14,15]. As expected, the Klein-Nishina cross section exhibits discrepancies with respect to experimental data at low energies; most of the scattering function options result in similar accuracy with respect to experimental data, with the exception of those calculated by Kahane.

Validation of Doppler broadening and of other features of Compton scattering implemented in Geant4 is in progress at the time of writing this extended abstract.

An example of comparison between simulation models and experimental data is shown in Figure 2.

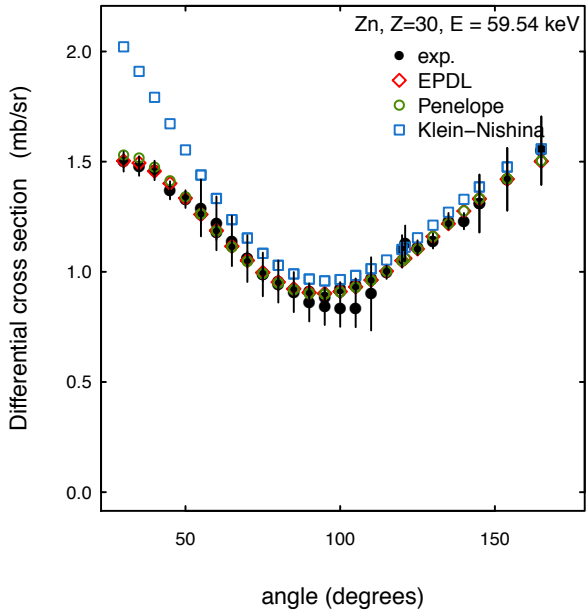

Fig. 2 An example of total Compton scattering differential cross section validation: experimental measurements (black symbols) and a selection of cross section calculations considered in the validation analysis.

**Photoionization**

The validation process evaluated cross section tabulations and parameterizations published by Biggs and Lighthill [16], Brennan and Cowan [17], Chantler [18,19], Ebel [20], Elam [21], Henke [22,23], McMaster [24,25], Scofield [26], Storm and Israel [11], Veigele [27] and Verner [28], as well as the tabulations in the EPDL97 [8], PHOTX [29], RTAB [4] and XCOM [30] data libraries. It encompassed both total and partial cross sections (i.e. for the ionization of a given shell). At the time of writing this extended abstract the validation concerned photon energies above 100 eV; the analysis is in progress in the lower energy range.

Categorical statistical tests show that most total cross section calculation methods are equivalent in terms of compatibility with experimental data, with the exception of those calculated by Chantler and by Brennan and Cowan. Some differences in accuracy are observed regarding the original cross section parameterization by Biggs and Lighthill and a modified version of it implemented in Geant4.

Similarly, most models of inner shell cross sections reproduce experimental data consistently with their uncertainties, with the exception of Ebel's parameterization for the K shell, which appears significantly less accurate.

Calculated outer shell photoionization cross sections are statistically incompatible with experimental data; nevertheless, due to the limited experimental data sample at this stage of the analysis one cannot draw firm conclusions regarding the accuracy of the models.

An example of comparison between simulation models and experimental data is shown in Figure 3.

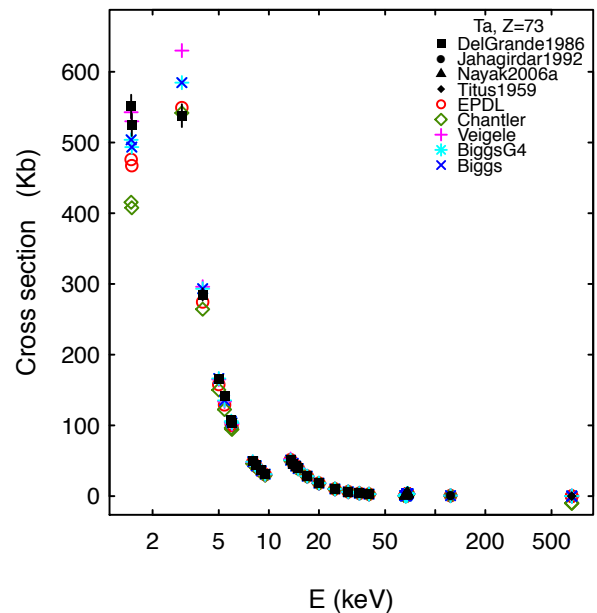

Fig. 3 An example of total photoionization cross section validation: experimental measurements (black symbols) and a selection of cross section calculations considered in the validation analysis.

**Photon conversion into $e^+e^-$ pairs**

At the present stage of the project total pair production cross sections tabulated in EPDL, XCOM, Penelope and parameterized in Geant4 standard electromagnetic package have been examined.

Different accuracy is observed associated with the various modeling options in the low energy range close to the threshold, despite they are all based on the same calculations of Hubbell et al. [10]. The validation process at higher energies is in progress at the time of writing this extended abstract. An example of comparison between simulation models and experimental data is shown in Figure 4.

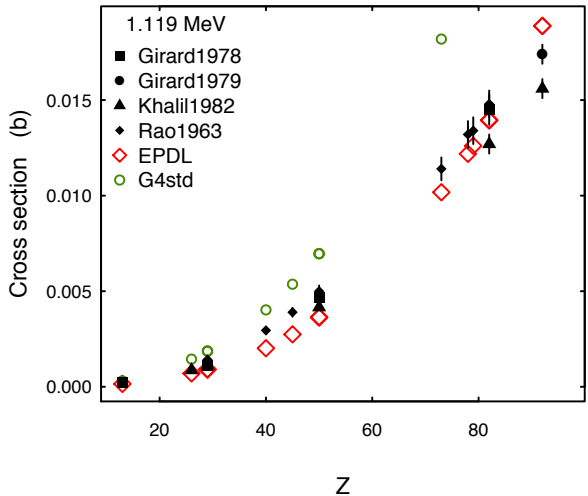

Fig. 4 An example of $e^+e^-$ pair production cross section validation: experimental measurements (black symbols) and a selection of cross section calculations considered in the validation analysis.

**Electron impact ionization**

A previously published evaluation [31] concerned total cross sections for electron impact ionization, with special emphasis on the low energy range below 1 keV: cross sections calculations based on EEDL [32], on the Binary-Encounter-Bethe [33] model and the Deutsch-Märk [34] model were compared to a wide collection of experimental data. The statistical analysis identified the Deutsch-Märk model as the one producing the most accurate cross sections.

Recently, the evaluation was extended to partial cross sections (i.e. cross sections for the ionization of a specific shell), also including the model by Bote and Salvat [35] for the calculation of inner shell cross sections.

An example of comparison between simulation models and experimental data is shown in Figure 5.

**SOFTWARE FEATURES**

All the physics models mentioned in this extended abstract are implemented according to a common software design, which is based on policy classes [36] and is compatible with Geant4 kernel. Models currently implemented in Geant4 are refactored [37].

The adopted software design enables a flexible configuration of physics processes and facilitates their test.

**CONCLUSION**

A project is in progress to thoroughly and quantitatively validate electromagnetic models suitable to be used with Geant4. The lightweight software design, minimal dependencies on other software and robust validation makes the package resulting from this project a relevant component to be considered for future evolutions of simulation software currently discussed at CERN.

**ACKNOWLEDGMENTS**

The support of the CERN Library has been essential for this research.

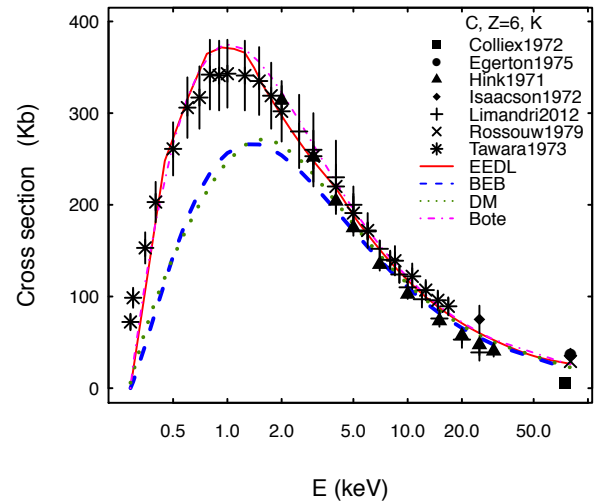

Fig. 5 An example of validation of K-shell ionization cross section by electron impact: experimental measurements (black symbols) and a selection of cross section calculations considered in the validation analysis.